# EFFECTS OF MISSING OBSERVATIONS ON PREDICTIVE CAPABILITY OF CENTRAL COMPOSITE DESIGNS

Yisa Yakubu[1*], Angela Unna Chukwu[2], Bamiduro Timothy Adebayo[2], Amahia Godwin Nwanzo[2]

[1]Department of Mathematics and Statistics, Federal University of Technology, Minna, Niger State, Nigeria

[2]Department of Statistics, University of Ibadan, Ibadan, Oyo state, Nigeria

*Abstract*

*Quite often in experimental work, many situations arise where some observations are lost or become unavailable due to some accidents or cost constraints. When there are missing observations, some desirable design properties like orthogonality, rotatability and optimality can be adversely affected. Some attention has been given, in literature, to investigating the prediction capability of response surface designs; however, little or no effort has been devoted to investigating same for such designs when some observations are missing. This work therefore investigates the impact of a single missing observation of the various design points: factorial, axial and center points, on the estimation and predictive capability of Central Composite Designs (CCDs). It was observed that for each of the designs considered, precision of model parameter estimates and the design prediction properties were adversely affected by the missing observations and that the largest loss in precision of parameters corresponds to a missing factorial point.*

## 1.Introduction

Central composite design (CCD) is the most popular class of second-order response surface designs, which was introduced by Box and Wilson (1951). This design consists of factors with five levels that involve three categories:

  i. complete (or a fraction of) $2^k$ factorial design with factor levels coded as -1, 1(called the factorial portion),
  ii. an axial portion consisting of $2k$ points arranged so that two points are chosen on the coordinate axis of each control variable at a distance of α from the design center,
  iii. $n_0$ center points.
  iv. Thus the total number of points in a CCD is $n = 2^k + 2k + n_0$. The second-order response surface model for these designs is

$$y = \beta_0 + \sum_{i=1}^{k} \beta_i x_i + \sum_{i=1}^{k} \beta_{ii} x_i^2 + \sum_{i=1}^{k-1} \sum_{j=i+1}^{k} \beta_{ij} x_i x_j + \varepsilon \qquad (1.1)$$

Where **y** is the response variable, **x** is the input variable, *β* is a model coefficient, and ***e*** is a random error component.

One of the functions of the fitted response surface model, and of course, the primary goal of many designed experiments, is to allow for good prediction of response values at various points of interest throughout the experimental region. In response surface methodology, interest is more on prediction than parameter estimation since the points on the fitted surface are predicted responses. Thus the prediction variance criteria are essential tools for selecting response surface designs as the researcher is provided with variance information regarding the worst prediction scenario. Box and Hunter (1957) noted that consideration of only the variances of the individual model coefficient estimates does not, for the case of second or higher-order models, lead to any unique class of "best" designs. Their argument is that the precision of the coefficient estimates should be studied simultaneously. Several measures of prediction performance exist for comparing designs of which the most commonly considered is the scaled prediction variance (SPV).

For the second-order response surface model in (Equation 1.1), the scaled prediction variance of the expected response is given by

$$v(x) = \frac{N \text{Var}[\hat{y}_{(x)}]}{\sigma^2} = N f'(x)(X'X)^{-1} f(x) \qquad (1.2)$$

where $N$ is the design size, $\sigma^2$ is the observation error, and $f(x)$ is the general form of the $1xp$ model vector.

Desirable designs are those with the smallest maximum SPV, and with reasonably stable SPV (i.e. smallest range) in the design region. The SPV allows the practitioner to measure the precision of the predicted response on a per observation basis and it penalizes larger designs over small designs.

Directly associated with the prediction variance $v(x)$ are the maximum (***G***) and the integrated (***V***) prediction variance optimality criteria. The ***G*-optimality** and the corresponding ***G*-efficiency** emphasize the use of designs for which the maximum scaled prediction variance $v(\mathbf{x})$ in the design region *is not too large*. Thus, a ***G*-optimal design** ($\xi$) is one that

$$\text{Min}_\zeta [\max_\zeta v(\mathbf{x})] \qquad (1.3)$$

The *G*-efficiency is determined as

$$G_{eff} = \frac{p}{\max_{x \in R} v(\mathbf{x})} \qquad (1.4)$$

The **V-criterion** attempts to generate a single measure of prediction performance through an average process, that is, $v(\mathbf{x})$ is averaged over some region of interest $R$ (Myers *et al.*, 2009 ). For a design $(\xi)$, the V-optimality criterion can be expressed as

$$V \to \underset{\xi}{Min} \frac{1}{K} \int_R v(x) dx$$

$$= \underset{\xi}{Min} \left\{ \frac{1}{K} \int_R f(x)'(X'X)^{-1} f(x) dx \right\} \qquad (1.5)$$

Since $f(x)'(X'X)^{-1}f(x)$ is a scalar, (1.5) can be written as

$$V = \underset{\xi}{Min} \frac{1}{K} \int_R tr[f(x)'(X'X)^{-1} f(x)] dx$$

$$= \underset{\xi}{Min} \frac{1}{K} \int_R tr[f(x)f(x)'(X'X)^{-1}] dx$$

$$= \underset{\xi}{Min} \left\{ tr \left\{ (X'X)^{-1} \left[ (1/K) \int_R f(x)f(x)' dx \right] \right\} \right\}$$

$$(1.6)$$

Where $K = \int_R dx$ is the volume of the region $R$, and the matrix $(1/K) \int_R f(x)f(x)' dx$ is a

$p \times p$ matrix of region moments.

The *A*-criterion is also often used to quantify the precision of parameter estimates, which is a way of assessing the design's capability to estimate the underlying model. The *A*-criterion minimizes the trace of the covariance matrix, that is, an *A-optimal* design is one in which we have

$$\min_\zeta trace(X'X))^{-1} \qquad (1.7)$$

There has been much discussion concerning the distance, α, of the axial points from the design center, in response surface designs. The choice of the axial distance (α) and the number of center runs ($n_c$) have substantial influence on the performance of CCDs. When the scaled prediction variance (SPV) is constant for any location of a fixed distance from the design center, the design is *rotatable* (Box and Hunter, 1957).

Missing observations adversely affect the orthogonality, rotatability, and optimality properties of a design. These properties are design performance criteria used for choosing "good" response surface designs. Extensive studies have been carried out in literature on predictive capability of full central composite designs both for spherical and cuboidal regions. For instance, Box and Draper (1959, 1963) proposed designs that are robust to model misspecification in terms of integrated average prediction variance (*V)*.

Ahmad and Gilmour (2010) study the robustness of subset response surface designs to a missing value in terms of prediction variance by computing the ratio of prediction variances for the design with a missing observation to the prediction variance for the full design. For each of the designs considered, the authors plotted the maximum and minimum ratios of variances against $1 <$ *radius* $< 1.8.$ They observed that the minimum ratio of prediction variances were quite robust

International Journal on Computational Sciences & Applications (IJCSA) Vol.4, No.6, December 2014to missing design points for almost all designs and for all types of missing design points except few. The authors also noted that the maximum ratio decreases gradually with the increase of radius for some designs with some type of missing points.

However, not much work has been done in literature on predictive capability of completely randomized central composite designs with some observations missing. This study therefore examines the predictive capability of standard central composite designs in the presence of missing observations.

For each of the CCDs we considered in this work, the effect of missing a single observation of the different design points on the precision of estimates of the model parameters is first examined, then we investigate the same effect on the maximum and average prediction variance criteria under different values of the axial point distance (α) from the design center.

## 2. Materials and Methods

Consider a second order response surface model in k variables and n design points as given in equation (1.1) above. Over the n design points, this model can be expressed in matrix form as

$$\underline{Y} = X\underline{\beta} + \underline{\varepsilon}$$

(2.1)

Where $\underline{Y}$ is the n x 1 response vector, $X$ is the n x p model matrix, $\underline{\beta}$ is the p x 1 vector of partameters, and $\underline{\varepsilon}$ is the n x 1 vector of random error terms.

When all the observations are available, the standard least squares estimate of the parameter $\underline{\beta}$ is given as

$$\hat{\underline{\beta}} = (X'X)^{-1}X'\underline{y}$$

The fitted model of the response variable $\underline{Y}$

$$\hat{\underline{Y}} = X\hat{\underline{\beta}} \qquad (2.2)$$

which is used to estimate the response variable at any point in the design region.
Now, the variance-covariance matrix of the estimator $\hat{\underline{\beta}}$ is

$$\text{Var}\left(\hat{\underline{\beta}}\right) = (X'X)^{-1}X'\text{Var}\left(\underline{y}\right)X(X'X)^{-1}$$

$$= (X'X)^{-1}X'\text{Var}(X\beta + \underline{\varepsilon})X(X'X)^{-1}$$

$$= \sigma_\varepsilon^2 (X'X)^{-1} \qquad (2.3)$$

The *A*-criterion aims to improve the precision of estimates of the model parameters by minimizing the trace of this variance-covariance matrix.

The variance of the predicted response is

$$V(\underline{\hat{Y}}) = XV(\underline{\hat{\beta}})X' = \sigma_\varepsilon^2 X(X'X)^{-1}X' = \sigma^2 H \qquad (2.4)$$

Where $H = X(X'X)^{-1}X'$ is the 'hat' matrix of the design with the property that $\text{trace}(H) = p$, $p$ being the number of model parameters.

If any m observations are missing, we partition the response vector $\underline{Y}$ and the model matrix X as

$$\begin{bmatrix} \underline{Y_m} \\ \cdots \\ \underline{Y_r} \end{bmatrix} \text{ and } \begin{bmatrix} \underline{X_m} \\ \cdots \\ \underline{X_r} \end{bmatrix}$$

respectively, where $\underline{Y_m}$ consists of m missing observations and $\underline{X_m}$ consists of m corresponding rows. Thus the information matrix can be expressed as

$$X'X = X'_m X_m + X'_r X_r$$

The ordinary least squares estimator of the parameters of the residual design is

$$\underline{\hat{\beta}^*} = (X'_r X_r)^{-1} X'_r \underline{y_r}$$

And the variance-covariance matrix of this estimator is

$$\text{Var}(\underline{\hat{\beta}^*}) = (X'_r X_r)^{-1}$$

The desire is always to minimize the variance of $\underline{\hat{\beta}}$ ($A - criterion$) so as to increase the precision of the parameter estimates, but whenever some observations are missing, the variance is increased (i.e., the estimates become less precise) as a result and such increase in the variance is given by

$$i.v. = \text{trace}(X'_r X_r)^{-1} - \text{trace}(X'X)^{-1} \qquad (2.5)$$

**2.1. Loss function in terms of the Trace Criterion**

Expression (2.5) measures the amount of reduction in the precision of the estimates when an observation is missing. Therefore, we define the **loss function** in terms of the $\text{trace}(A)$ criterion when some observations are missing as the relative reduction in the precision of the parameter estimates. This is given by

$$r.i.v. = \frac{\text{trace}(X'_r V_r^{-1} X_r)^{-1}}{\text{trace}(X'V^{-1}X)^{-1}} - 1 \qquad (2.6)$$

This expression is used in this work to estimate the loss in precision of estimates of the model parameters when some observations are missing.

## 2.2 Efficiency in terms of Maximum and Average Prediction Variances Criteria

In order to examine the effects of missing observations on the predictive capability of response surface designs, efficiency of each of the resulting residual design was examined relative to the full design in terms of these criteria and thus we compute the relative efficiencies

$$RE_G = \frac{MAX_{X \epsilon R} v(x)}{MAX_{X \epsilon R} v(x)_{reduced}} \quad (2.7)$$

and

$$RE_V = \frac{trace\left\{(X'X)^{-1}\left[\frac{1}{K}\int_R f(x)f'(x)dx\right]\right\}}{trace\left\{(X'X)^{-1}\left[\frac{1}{K}\int_R f(x)f'(x)dx\right]\right\}_{reduced}} \quad (2.8)$$

Where $f(x)$, $v(x)$, and $\frac{1}{K}\int_R f(x)f'(x)dx$ are as defined above.

And we note that

(i) Relative efficiency greater than 1 indicates that the missing observation has little or no adverse effect on the design in terms of this criterion, which means that the criterion is quite robust to the missing point.

(ii) Relative efficiency smaller than 1 indicates that the missing point has large adverse effect on the design in terms of the criterion.

Four different central composite designs were considered in this work. These designs are given in the table below, where $k$ is the number of design variables, $n_f$ is the number of factorial points, $n_\alpha$ is the number of axial points, $n_0$ is the number of center points, and $n$ is the total number of design points.

**Table 2.1**: Candidate CCDs considered.

| $k$ | $n_f$ | $n_\alpha$ | $n_0$ | $n$ |
|---|---|---|---|---|
| 2 | 4 | 4 | 4 | 12 |
| 3 | 8 | 6 | 4 | 18 |
| 4 | 16 | 8 | 4 | 28 |
| 5 | 32 | 10 | 4 | 46 |

## 3.Results and Discussions

In this section, the losses in precision of parameter estimates, the maximum and average prediction variances computed under various α values for full and residual designs were presented. The plots of the losses and that of the relative efficiencies were also presented.

TABLES 1a, 2a, 3a, and 4a below consist of the losses in precision of the model parameter estimates for the respective CCDs with one observation missing. The corresponding loss curves were given by FIGURES 1a, 2a, 3a, and 4a. We can see from these tables that the *trace*-criterion values for the full designs decreases as the distance of the axial points from the design center increases. We observed that missing observations of each of the design point categories (factorial, axial, and center) has adverse effect on the parameters.

We can observe from TABLES 1a and 2a and also from the corresponding figures (1a and 2a) that the loss due to missing a factorial point was the highest at $\alpha = 1.00$. In FIGURE 1a, the precision continues to improve sharply as $\alpha$ increases and becomes stable at $\alpha > 1.41$, while in 2a, it decreases as $\alpha$ increases up to 1.21 after which it improves and becomes slightly stable at $\alpha > 1.68$. As we can see from FIGURES 3a and 4a, this criterion becomes slightly robust to a missing factorial point as the number of design variables increases.

Also from Tables and Figures 1a and 2a, we observed that the loss in precision due to a missing axial point was the second largest. From Figure 1a, we observed a slight improvement in precision as $\alpha$ exceeds 1.00 and the precision becomes stable for $1.21 < \alpha < 1.50$ after which it decreases sharply for the rest $\alpha$ values. From Figure 2a, the precision improves sharply as $\alpha$ increases and a sharp loss was again observed as $\alpha$ goes beyond 1.68. From Figures 3a and 4a, missing axial point causes the highest loss in precision of parameter estimates at $\alpha = 1.00$ as we can directly observe. However, the loss in precision due to this missing (axial) point continues to decrease sharply as $\alpha$ increases to 2.00 in Figure 3a and to 2.24 in Figure 4a after which the loss increases for the rest $\alpha$ values in both cases.

We observed from Figure 1a that the loss due to the missing center point is the lowest and makes a bell-shaped curve with the $\alpha$ values while in Figures 2a, 3a and 4a, the bell shape increases as the number of design variables increases. In Figure 2a, the loss curve due to this missing (center) point intersects that due to the missing axial point at two distance points and has its peak at $\alpha \cong 1.71$. In Figures 3a and 4a, this loss curve intersects the other two curves and has its peak at $\alpha \cong 2.11$ and $\alpha \cong 2.31$ respectively.

In general, the loss in precision of parameter estimates due to a missing factorial point continues to decrease as the number of design factors increases.

TABLES 1b, 2b, 3b and 4b give the values of the maximum and average prediction variances for the complete designs and for designs with a single observation missing. These tables show that the location of the maximum prediction variance is restricted to the factorial portion of the designs at low values of α and to the axial portion as α increases.

Figures 1b, 2b, 3b, and 4b give the relative *G*-efficiency plots respectively for the designs considered here. We can observe from each of these figures that the *G*-criterion is quite robust to the missing center point. In figure 1b, efficiency loss was only due to the missing axial point at low values of α after which it continues to decrease sharply. In figure 2b, the largest loss in efficiency corresponds to the missing factorial point and this loss remains slightly stable for $1.00 \leq \alpha \leq 1.732$, after which it continues to decrease as α increases. In this figure also, loss in efficiency due to the missing axial point was observed as α approaches 1.732, which continues to increase and then decreases slowly as α exceeds 2.00. In figure 3b, we observed a loss in efficiency corresponding to the missing axial point and also to the missing factorial point. The loss due to the missing factorial point increases gradually at low values of α and then suddenly decreases as α increases beyond 2.00. The loss corresponding to the missing axial point is slightly stable as α goes beyond 2.00. In figure 4b, we observe the largest loss in efficiency corresponding to the missing axial point at α = 1.00, which reduces gradually as α increases, and as α exceeds 2.236, the loss becomes slightly robust to changes in α values. We also observed the smallest loss in efficiency from this figure, which corresponds to the missing factorial point at α = 1.00, which suddenly decreases as α increases.

Figures 1c, 2c, 3c, and 4c give the respective *V*-efficiency plots for the designs under discussion. We observed the largest efficiency loss corresponding to missing a factorial point in figures 1c and 2c, which continues to decrease as α increases. In figure 1c, we observed a slightly stable efficiency loss corresponding to the missing axial point for $1.00 \leq \alpha \leq 1.50$, after which it starts to increase, while that of figure 2c makes a bell-shaped curve with α values. The loss due to missing axial point in figure 3c and 4c also decreases up to a certain value of α and then starts to increase again. The loss in efficiency due to missing center point is only observed in figure 1c as α exceeds 1.10 and increases gradually as α increases up to 1.414 after which it becomes stable as α continues to increase. The smallest efficiency loss is recorded for the missing center point at α = 1.210 in figure 2c, which gradually increases as α increases and then starts to decreases again. In figure 3c, we observed the smallest efficiency loss due to the missing center point as α exceeds 1.210; the same loss was observed in figure 4c as α exceeds 1.00, which continues to increase as α increases up to the point $\alpha \approx 2.11$ in figure 3c and $\alpha \approx 2.25$ in figure 4c, and then decreases again.

### 3.1 Two-factor CCD

This design consists of $n_f$ = 4 factorial points, $n_\alpha$ = 4 axial points, $n_0$ = 4 center points, and thus N = 12 total design points.

**Table 1a**: Loss in precision of parameter estimates due to single missing observations in 2-variable CCD

| Axial point distance | *A*-criterion value for full design | Loss in precision (*lp*) due to missing | | |
|---|---|---|---|---|
| | | factorial(*f*) | axial($\alpha$) | center(*c*) |
| 1.000 | 1.5416 | 0.4702906 | 0.2072522 | 0.06117 |
| 1.210 | 1.2440 | 0.3397106 | 0.1685691 | 0.098553 |
| 1.414 | 1.0626 | 0.2550348 | 0.176454 | 0.117636 |
| 1.500 | 0.9967 | 0.2362797 | 0.1902278 | 0.118491 |
| 2.000 | 0.7187 | 0.2319466 | 0.3015166 | 0.090024 |

**Table 1b**: Scaled prediction Variance and average prediction variance for complete design and for designs with one observation missing

No. of variables k = 2. Total design points n = 12
No. of parameters p = 6. No. of centre points = 4

| $\alpha$ | Missing point | $v(x)$ at | | | *V* |
|---|---|---|---|---|---|
| | | Factorial | Axial | Center | |

| | | point | point | point | |
|---|---|---|---|---|---|
| 1.000 | None | **9.500** | 6.000 | 2.500 | 3.633 |
| | Factorial | **9.533** | 5.866 | 2.383 | **4.913** |
| | Axial | **8.861** | 6.111 | 2.444 | 3.931 |
| | Center | **8.732** | 5.596 | 2.894 | 3.586 |
| 1.21 | None | **8.464** | 6.669 | 2.866 | 3.318 |
| | Factorial | **8.370** | 6.252 | 2.661 | **3.930** |
| | Axial | **9.819** | 6.712 | 2.670 | 3.535 |
| | Center | **7.772** | 6.139 | 3.451 | 3.410 |
| 1.414 | None | **7.500** | 7.499 | 2.999 | 3.166 |
| | Factorial | **7.334** | 6.952 | 2.749 | **3.483** |
| | Axial | 6.954 | **7.332** | 2.749 | 3.361 |
| | Center | **6.875** | 6.8741 | 3.666 | 3.351 |
| 1.5 | None | 7.159 | **7.861** | 2.979 | 3.109 |
| | Factorial | 6.995 | **7.283** | 2.737 | **3.364** |
| | Axial | 6.652 | **7.710** | 2.737 | 3.311 |
| | Center | 6.566 | **7.209** | 3.633 | 3.312 |
| 2.00 | None | 6.000 | **9.500** | 2.500 | 2.766 |
| | Factorial | 6.111 | **8.861** | 2.444 | 2.943 |
| | Axial | 5.866 | **9.533** | 2.383 | **3.098** |
| | Center | 5.596 | **8.732** | 2.894 | 2.931 |

## 3.2. Three-factor CCD

This design consists of $n_f = 8$ factorial points, $n_\alpha = 6$ axial points, and $n_0 = 4$ center points, with $N = 18$ total design points.

**Table 2a**: Loss in precision of parameter estimates due to single missing observations in 3-variable CCD

| Axial point distance | $A$-criterion value for full design | Loss in precision ($lp$) due to missing | | |
|---|---|---|---|---|
| | | factorial($f$) | axial($\alpha$) | center($c$) |
| 1.000 | 1.9369 | 0.2111622 | 0.1953637 | 0.021116 |
| 1.210 | 1.4227 | 0.233078 | 0.1447951 | 0.044071 |
| 1.681 | 1.0814 | 0.1827261 | 0.088034 | 0.103292 |
| 1.732 | 1.0575 | 0.1780615 | 0.088227 | 0.105059 |
| 2.000 | 0.9333 | 0.1692918 | 0.1031823 | 0.094718 |
| 2.250 | 0.8322 | 0.1760394 | 0.1237683 | 0.07366 |
| 2.500 | 0.7549 | 0.1855875 | 0.1413432 | 0.056431 |
| 3.000 | 0.6572 | 0.1982654 | 0.1653987 | 0.037279 |

**Table 2b**: Scaled prediction Variance and average prediction variance for complete design and for designs with a single observation missing

No. of variables k = 3 Total design points n = 18
No. of parameters p = 10 No. of center points = 4

| Axial point distance ($\alpha$) | Missing point | $v(x)$ at | | | IV |
|---|---|---|---|---|---|
| | | Factorial point | Axial point | Center point | |
| 1.000 | None | **14.292** | 9.085 | 2.785 | 5.607 |
| | Factorial | **16.606** | 9.060 | 2.677 | **6.647** |
| | Axial | **13.698** | 11.769 | 2.942 | 6.098 |
| | Center | **13.510** | 8.763 | 3.112 | 5.497 |
| 1.21 | None | **13.685** | 9.502 | 3.374 | 4.817 |
| | Factorial | **16.091** | 9.341 | 3.249 | **5.575** |
| | Axial | **13.111** | 11.409 | 3.425 | 5.110 |
| | Center | **12.943** | 9.113 | 3.922 | 4.855 |
| rotatable (1.681) | None | **12.059** | 10.929 | 4.486 | 4.543 |
| | Factorial | **14.124** | 10.465 | 4.239 | 4.868 |
| | Axial | 11.509 | **11.938** | 4.240 | 4.632 |
| | Center | **11.390** | 10.324 | 5.643 | **4.944** |
| 1.732 | None | **11.893** | 11.142 | 4.499 | 4.527 |
| | Factorial | **13.932** | 10.660 | 4.249 | 4.828 |
| | Axial | 11.354 | **12.142** | 4.249 | 4.609 |
| | Center | **11.285** | 10.562 | 5.663 | **4.948** |
| 2.00 | None | 11.175 | **12.300** | 4.200 | 4.344 |
| | Factorial | **13.263** | 11.769 | 4.016 | 4.590 |
| | Axial | 10.736 | **13.421** | 4.026 | 4.448 |
| | Center | 10.578 | **11.641** | 5.173 | **4.745** |
| 2.25 | None | 10.729 | **13.247** | 3.670 | 4.078 |
| | Factorial | **13.089** | 12.713 | 3.601 | 4.332 |
| | Axial | 10.442 | **14.536** | 3.594 | 4.247 |
| | Center | 10.201 | **12.554** | 4.354 | **4.377** |
| 2.50 | None | 10.416 | **13.989** | 3.183 | 3.811 |
| | Factorial | 13.138 | **13.461** | 3.205 | **4.081** |
| | Axial | 10.309 | **15.350** | 3.160 | 4.042 |
| | Center | 9.940 | **13.253** | 3.652 | 4.018 |
| 3.00 | None | 9.972 | **15.012** | 2.536 | 3.392 |
| | Factorial | 13.401 | **14.474** | 2.650 | 3.678 |
| | Axial | 10.213 | **16.235** | 2.530 | **3.713** |
| | Center | 9.550 | **14.204** | 2.788 | 3.497 |

### 3.3. Four-factor CCD

This design consists of $n_f$ = 16 factorial points, $n_\alpha$ = 8 axial points, and $n_0$ = 4 center points, with $n = 28$ total design points.

**Table 3a**: Loss in precision of parameter estimates due to single missing observations in 4-variable CCD

| Axial point distance | $A$-criterion value for full design | Loss in precision ($lp$) due to missing | | |
|---|---|---|---|---|
| | | factorial($f$) | axial($\alpha$) | center($c$) |
| 1.000 | 2.2675 | 0.0480706 | 0.1753032 | 0.009261 |
| 1.210 | 1.4871 | 0.0689933 | 0.1374487 | 0.020846 |

| 2.000 | 0.9583 | 0.0771157 | 0.047793 | 0.108734 |
| 2.250 | 0.8747 | 0.0784269 | 0.0533897 | 0.09649 |
| 2.500 | 0.7866 | 0.0845411 | 0.0652174 | 0.069286 |
| 3.000 | 0.6602 | 0.0960315 | 0.0802787 | 0.033475 |

**Table 3b**: Scaled prediction Variance and average prediction variance for complete design and for designs with a single observation missing

No. of variables k = 4 Total design points n = 28
No. of parameters p = 15 No. of center points = 4

| ($\alpha$) | Missing point | $v(x)$ at | | | IV |
|---|---|---|---|---|---|
| | | Factorial point | Axial point | Center point | |
| 1.000 | None | **18.446** | 13.932 | 3.347 | 7.477 |
| | factorial | **21.424** | 13.619 | 3.237 | 7.559 |
| | Axial | 17.913 | **21.461** | 3.634 | **8.365** |
| | Center | **17.791** | 13.666 | 3.666 | 7.400 |
| 1.21 | None | **18.104** | 14.294 | 3.992 | 5.628 |
| | factorial | **21.289** | 13.988 | 3.868 | 5.728 |
| | Axial | 17.605 | **20.769** | 4.245 | **6.140** |
| | Center | **17.465** | 14.010 | 4.489 | 5.699 |
| rotatable (2.00) | None | **16.333** | 16.333 | 7.000 | 5.211 |
| | factorial | **19.800** | 15.862 | 6.750 | 5.190 |
| | Axial | 15.862 | **19.800** | 6.750 | 5.257 |
| | Center | **15.750** | 15.750 | 9.000 | **6.075** |
| 2.25 | None | 15.815 | **17.623** | 6.491 | 4.983 |
| | factorial | **19.400** | 17.122 | 6.288 | 4.971 |
| | Axial | 15.400 | **21.255** | 6.344 | 5.048 |
| | Center | 15.266 | **17.035** | 8.148 | **5.759** |
| 2.50 | None | 15.431 | **18.913** | 5.447 | 4.512 |
| | factorial | **19.278** | 18.410 | 5.342 | 4.535 |
| | Axial | 15.118 | **22.907** | 5.455 | 4.651 |
| | Center | 14.929 | **18.319** | 6.521 | **5.050** |
| 3.00 | None | 14.878 | **20.885** | 3.713 | 3.613 |
| | factorial | 19.398 | **20.399** | 3.741 | 3.684 |
| | Axial | 14.825 | **25.121** | 3.815 | 3.855 |
| | Center | 14.434 | **20.208** | 4.128 | **3.838** |

### 3.4. Five - factor CCD

A full replicate of this design consists of $n_f = 32$ factorial points, $n_\alpha = 10$ axial points, and $n_0 = 4$ center points, i.e., there are a total of $n = 46$ design points.

**Table 4a**: Loss in precision of parameter estimates due to single missing observations in 5-variable CCD

| Axial point distance | $A$-criterion value for full design | Loss in precision ($lp$) due to missing | | |
|---|---|---|---|---|
| | | factorial($f$) | axial($\alpha$) | center($c$) |
| 1.000 | 2.5839 | 0.0109137 | 0.1580557 | 0.004915 |
| 1.500 | 1.0301 | 0.0260169 | 0.0940685 | 0.029803 |
| 2.236 | 0.8163 | 0.0285434 | 0.0292785 | 0.122504 |
| 2.378 | 0.7828 | 0.029254 | 0.0297649 | 0.117782 |
| 2.500 | 0.7460 | 0.0302949 | 0.033378 | 0.104155 |
| 2.750 | 0.6646 | 0.0341559 | 0.0430334 | 0.068914 |
| 3.000 | 0.5963 | 0.0379004 | 0.0489686 | 0.042428 |

**Table 4b**: Scaled prediction Variance and average prediction variance for complete design and for designs with a single observation missing

No. of variables k = 5 Total design points n = 46
No. of parameters p = 21 No. of center points = 4

| ($\alpha$) | Missing point | $v(x)$ at | | | IV |
|---|---|---|---|---|---|
| | | Factorial point | Axial point | Center point | |
| 1.000 | None | 22.570 | **22.592** | 4.456 | 12.587 |
| | factorial | **25.492** | 22.165 | 4.362 | 12.447 |
| | Axial | 22.144 | **38.628** | 4.878 | **14.157** |
| | Center | 22.080 | **22.393** | 4.827 | 12.520 |
| 1.500 | None | 22.063 | **23.307** | 6.720 | 6.801 |
| | factorial | 25.242 | **22.897** | 6.583 | 6.764 |
| | Axial | 21.685 | **36.209** | 7.095 | **7.288** |
| | Center | 21.590 | **23.101** | 7.698 | 7.140 |
| 2.236 | None | 20.946 | **24.971** | 11.499 | 7.609 |
| | factorial | 24.391 | **24.499** | 11.249 | 7.527 |
| | Axial | 20.576 | **33.571** | 11.249 | 7.614 |
| | Center | 20.491 | **24.428** | 14.999 | **9.177** |
| rotatable (2.378) | None | 20.724 | **25.817** | 11.158 | 7.468 |
| | factorial | 24.225 | **25.331** | 10.9208 | 7.392 |
| | Axial | 20.369 | **34.493** | 10.968 | 7.475 |
| | Center | 20.278 | **25.286** | 14.411 | **8.969** |
| 2.500 | None | 20.554 | **26.663** | 10.407 | 7.152 |
| | factorial | 24.120 | **26.169** | 10.199 | 7.089 |
| | Axial | 20.219 | **35.637** | 10.335 | 7.197 |
| | Center | 20.120 | **26.167** | 13.158 | **8.454** |
| 2.75 | None | 20.259 | **28.442** | 8.315 | 6.223 |
| | factorial | 24.008 | **27.942** | 8.187 | 6.195 |
| | Axial | 19.993 | **38.248** | 8.479 | 6.371 |
| | Center | 19.855 | **27.984** | 9.930 | **7.026** |
| 3.00 | None | 20.009 | **30.006** | 6.406 | 5.314 |
| | factorial | 23.972 | **29.511** | 6.344 | 5.315 |
| | Axial | 19.830 | **40.414** | 6.664 | 5.524 |
| | Center | 19.625 | **29.514** | 7.281 | **5.767** |

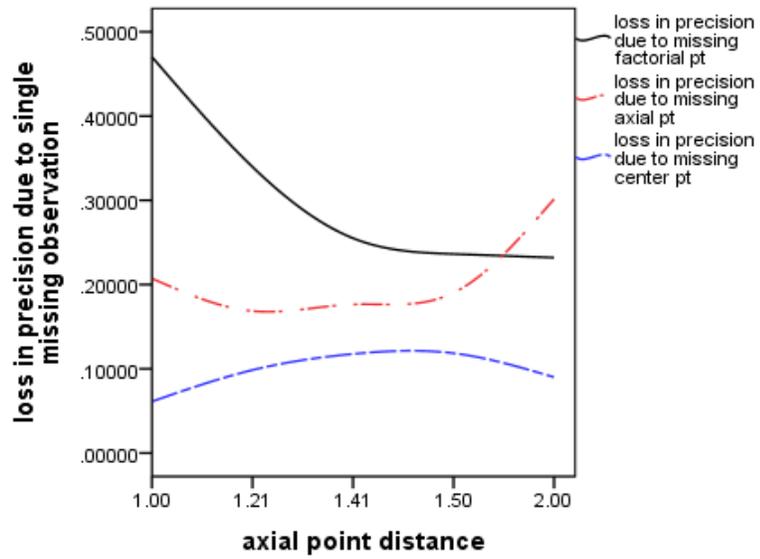

(a)

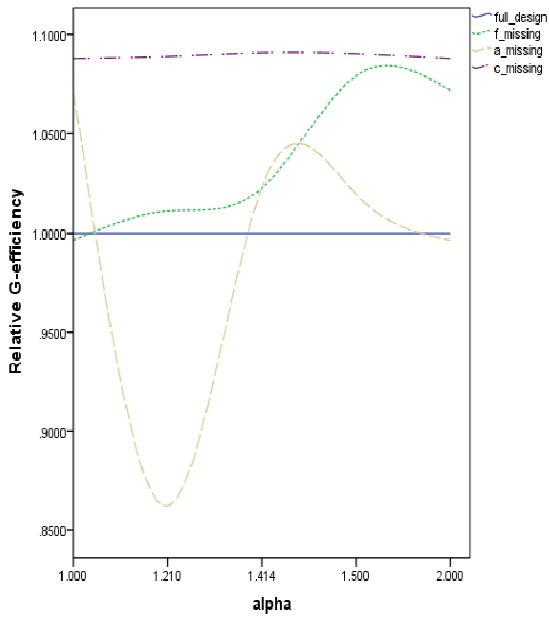

(b)

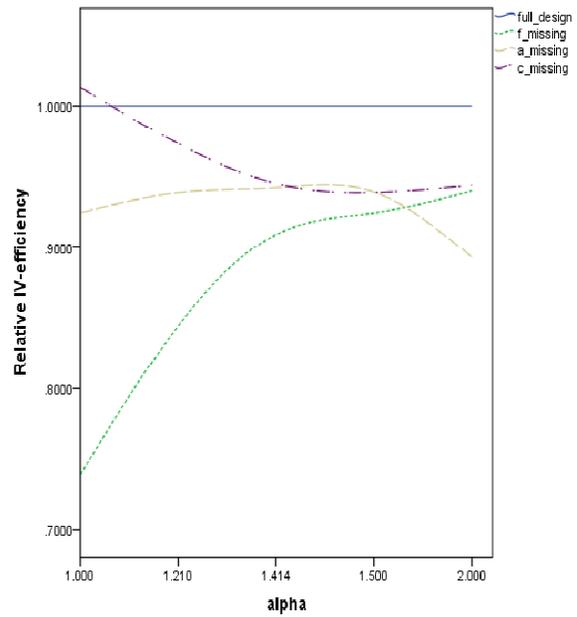

(c)

**Fig.1**: Loss curves due to single missing observation (a), and relative *G*, and *V*-efficiency plots for a 2-factor CCD under various values of α.

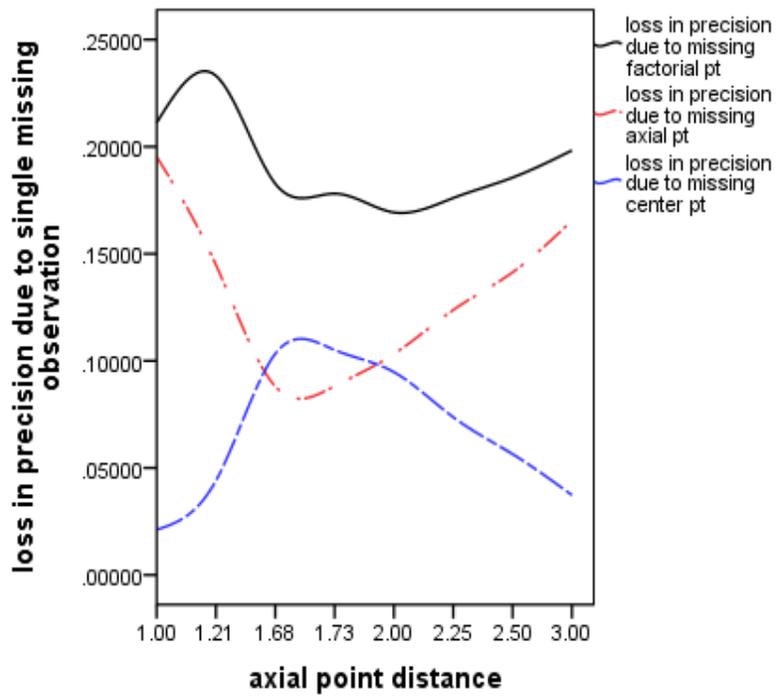

(a)

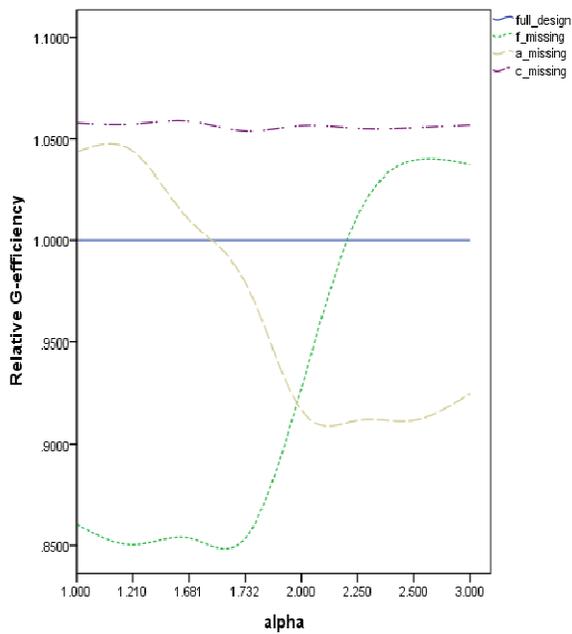

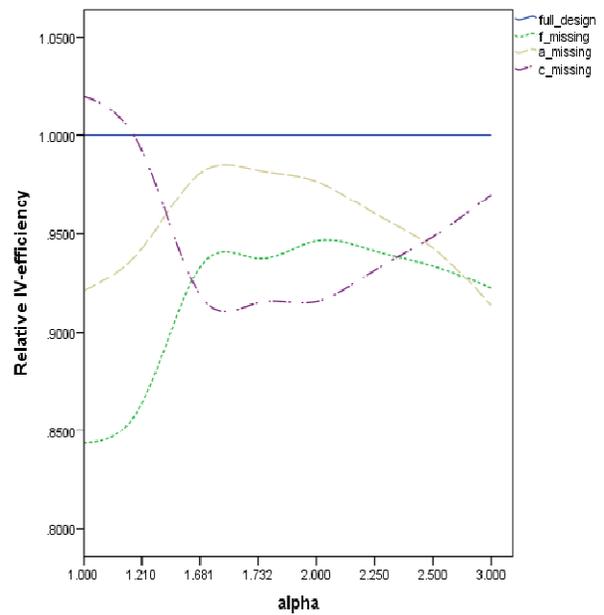

**(b)**        **(c)**

**Fig.2**: Loss curves due to single missing observation (a), and relative $G$, and $V$-efficiency plots ((b) and (c)) for a 3-factor CCD under various values of α.

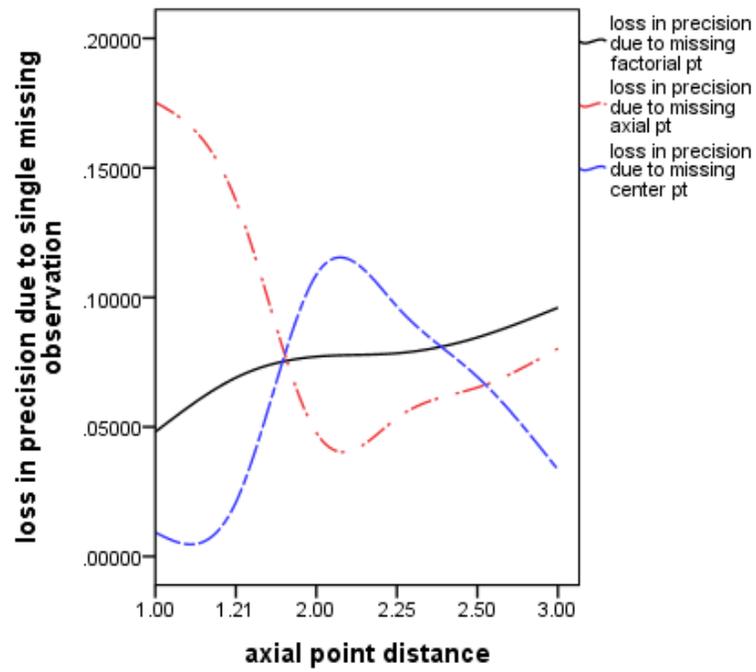

(a)

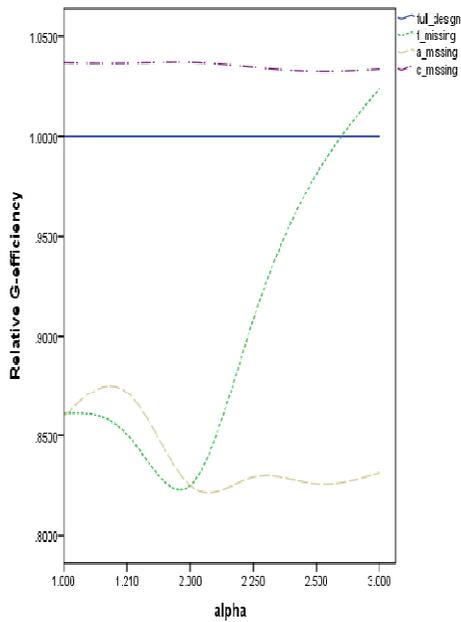

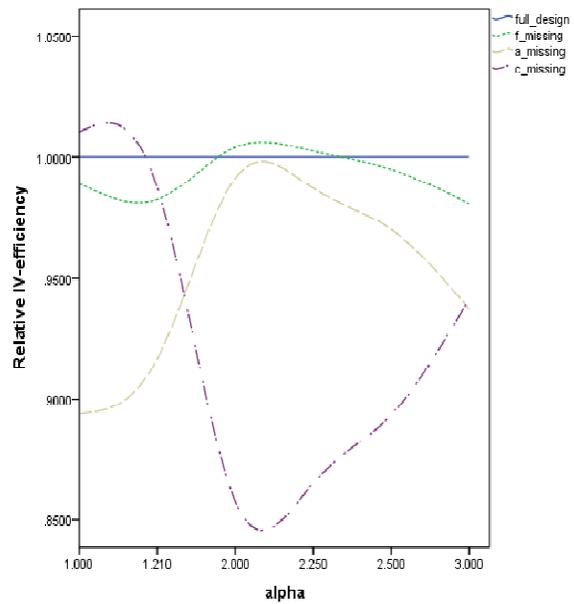

(b) (c)

**Fig.3**: Loss curves due to single missing observation (a), and relative *G*, and *V*-efficiency plots ((b) and (c)) for a 4-factor CCD under various values of α.

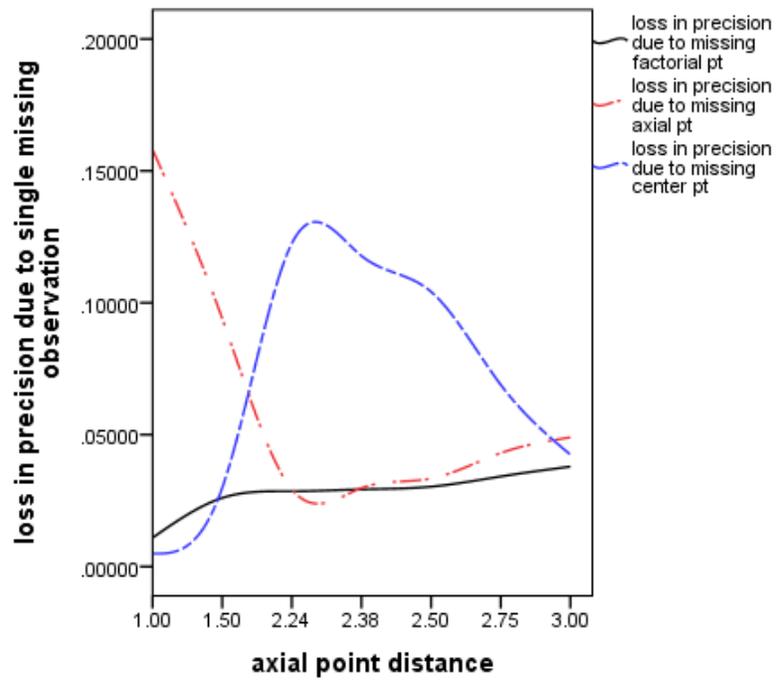

(a)

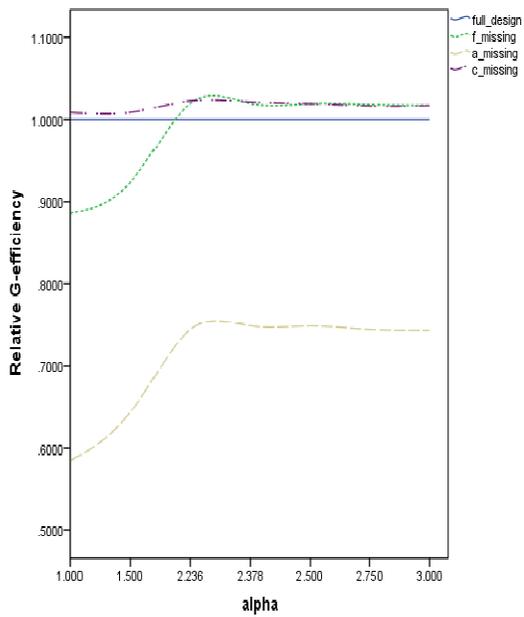

(b)

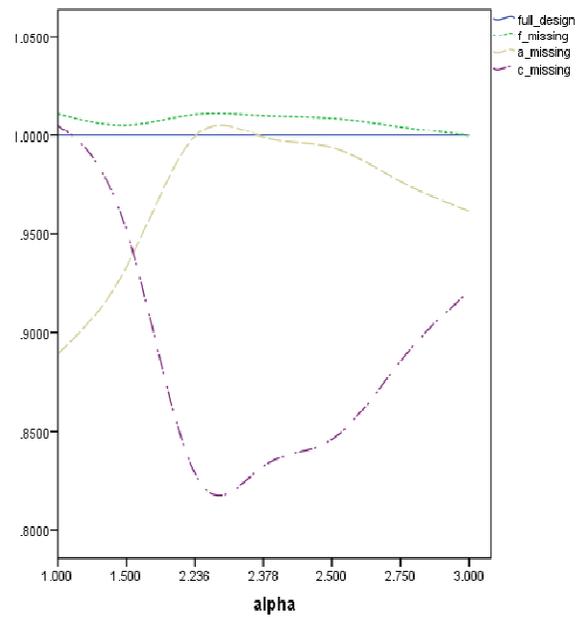

(c)

**Fig.4**: Loss curves due to single missing observation (a), and relative *G*, and *V*-efficiency plots ((b) and (c)) for a 5-factor CCD under various values of α.

International Journal on Computational Sciences & Applications (IJCSA) Vol.4, No.6, December 2014

## 4.Conclusions

This study has shown that missing a single observation of any of the three different categories of points in a central composite design (factorial, axial, and center points) adversely affects the designs predictive and estimation capabilities. The trace criterion is a decreasing function of α as can be seen directly from the presented tables; thus we can obtain a good design in terms of this criterion by taking large values for α during experimentation. The tables of variances of parameter estimates corresponding to the missing observations show adverse effects on the precision of the parameters (intercept, linear, quadratic, and interaction) due to these missing points. It was observed that the extent of the effect depends on the type of the missing point and also on the distance ($\alpha$) of the axial points from the design center. It was also shown in the tables of the maximum and the average prediction variances that the location of maximum lies only in the factorial portion of the designs for low values of α and then in the axial portion as α increases. The average prediction variances vary with the type of missing point and also with the value of α.

It was observed that for each of the four designs under discussion, the *G*-efficiency is robust to a missing center point but is highly affected by a missing axial point as the number of factors of the design increases. We also observed that the effect of missing a single factorial point on the design's predictive capability decreases as the number of factors of the design increases. For each of the designs, it was observed that the smallest loss in *A*-efficiency corresponds to a missing center point and at α = 1.00. Therefore, we recommend here that the practitioner should always endeavor to investigate the effect of missing observations on CCDs in terms of more than just one optimality criterion when faced with the challenges of the design choice. Knowledge of all these will alert the practitioner prior to data collection as to where in the design region he/she should expect the worst prediction and where to collect more data if necessary.

**Authors**

**Mr Yisa Yakubu** holds an M.Sc. degree in statistics, from the University of Ibadan, Nigeria and is currently a Ph.D. student in the Department of Statistics, University of badan, Oyo State, Nigeria. His research interests are experimental design and response surface methodology.

**Dr. Angela U. Chukwu,** is a Senior Lecturer in the Department of Statistics,University of Ibadan, Oyo state, Nigeria. She holds a Ph.D. degree in statistics from the University of Ibadan. Her research interests are biometrics and environmental statistics. 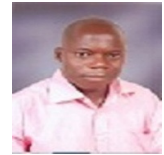